\documentclass[aps,prl,twocolumn,groupedaddress,showpacs,amsmath,superscriptaddress]{revtex4}
\usepackage{graphicx}

\begin{document}

\title{\boldmath High-Precision Measurement of the Laser-Trapping Frequencies for $^{209,210,211}$Fr Atoms}

\author{S.~Sanguinetti}
\email[Electronic address: ]{stefano.sanguinetti@df.unipi.it}
\affiliation{CNISM - Unit\`a di Pisa and Dipartimento di Fisica E.
Fermi, Universit\`a di Pisa, Largo B. Pontecorvo 3, 56127 Pisa,
Italy}
\author{R.~Calabrese}
\affiliation{Dipartimento di Fisica dell'Universit\`a degli Studi
and Istituto Nazionale di Fisica Nucleare, Sezione di Ferrara, Via
Saragat 1, 44100 Ferrara, Italy}
\author{L.~Corradi}
\author{A.~Dainelli}
\affiliation{Istituto Nazionale di Fisica Nucleare, Laboratori
Nazionali di Legnaro, Viale dell'Universit\`a 2, 35020 Legnaro
(PD), Italy}
\author{A.~Khanbekyan}
\author{E.~Mariotti}
\author{C.~de~Mauro}
\affiliation{CNISM - Unit\`a di Siena and Dipartimento di Fisica,
Universit\`a degli Studi di Siena, Via Roma 56, 53100 Siena,
Italy}
\author{P.~Minguzzi}
\affiliation{CNISM - Unit\`a di Pisa and Dipartimento di Fisica E.
Fermi, Universit\`a di Pisa, Largo B. Pontecorvo 3, 56127 Pisa,
Italy}
\author{L.~Moi}
\affiliation{CNISM - Unit\`a di Siena and Dipartimento di Fisica,
Universit\`a degli Studi di Siena, Via Roma 56, 53100 Siena,
Italy}
\author{G.~Stancari}
\author{L.~Tomassetti}
\affiliation{Dipartimento di Fisica dell'Universit\`a degli Studi
and Istituto Nazionale di Fisica Nucleare, Sezione di Ferrara, Via
Saragat 1, 44100 Ferrara, Italy}
\author{S.~Veronesi}
\affiliation{CNISM - Unit\`a di Siena and Dipartimento di
Fisica, Universit\`a degli Studi di Siena, Via Roma 56, 53100
Siena, Italy}

\date{\today}

\begin{abstract}
We present the accurate measurement of the frequency of the
$7S-7P$ laser-trapping transition for three francium isotopes. Our
approach is based on an interferometric comparison to deduce the
unknown laser frequency from a secondary laser frequency-standard.
After careful investigation of systematics, with samples of about
100 atoms the final accuracy reaches 8~MHz, an order of magnitude
better than the best previous measurement for $^{210}$Fr, and
opens the way to improved tests of the theoretical computation of
Fr atomic structure.
\end{abstract}



\maketitle

Francium, the heaviest alkali, is a promising atom for probing
fundamental symmetries in physics~\cite{flambaum}: there are
projects to measure parity violation in electron-nucleon
interactions (parameterized by weak charges) and in
nucleon-nucleon forces (via anapole moments), and also to search
for an electric dipole moment of the electron (EDM). All the
isotopes of Fr being radioactive make this research field
extremely challenging, due to the small quantities of atoms
available. Laser trapping is an effective tool that allows to
concentrate the samples in a small volume. This technique was
successfully employed to trap Na, K, Rb and Fr radioactive
isotopes, in view of measurements of fundamental interactions:
parity violation in nuclear beta decay of Rb~\cite{crane} and
$\beta - \nu$ correlation in Na, K~\cite{betanu1,betanu2}.
Wieman's group~\cite{wieman} performed spectroscopic studies on
trapped $^{221}$Fr coming from radioactive decay of $^{229}$Th.
The group at Stony Brook~\cite{orozco} produced and trapped Fr
online with their accelerator: an impressive amount of
spectroscopic information about wavelengths and lifetimes was
achieved, mainly for $^{210}$Fr.

The availability of high quality experimental data stimulated the
improvement of theoretical studies on the atomic structure of Fr
and its isoelectronic sequence: an example with extensive
reference to the relevant literature is Ref.~\cite{safronova}. The
success in trapping small samples of Fr also triggered the
proposal of a new approach to measure parity violation in
atoms~\cite{bouchiat}. Here we report on the measurement of
transition frequencies with a final accuracy an order of magnitude
better than the best one made so far. This will lead to a more
accurate comparison between experiment and the theory of Fr energy
levels, which constitutes a necessary step to properly investigate
fundamental symmetries.

Our experiment to produce and collect Fr in a magneto-optical trap
(MOT) was set up in Legnaro: the apparatus is described in
Ref.~\cite{stancari}. In short, Fr ions are produced by nuclear
fusion~\cite{corradi} of a 100~MeV $^{18}$O beam colliding on a
heated gold target; the ions are then injected in a secondary
electrostatic beam line, pass through a low-resolution mass
selector and are conveyed to a heated yttrium neutralizer inside
the pyrex MOT cell, where the atoms are captured by the lasers.
The cold cloud is imaged on a high sensitivity CCD camera, able to
detect as few as ten trapped atoms. We mainly trap the most
abundant isotope ($^{210}$Fr), but we also succeeded in trapping
the $^{209}$Fr and $^{211}$Fr isotopes.

We use a standard configuration for the MOT, with a tunable Ti:Sa
laser for trapping, a diode laser for repumping and a magnetic
field gradient of about 7~gauss/cm. For each isotope, the
repumping laser is tuned to the $7S(F=I-1/2)-7P_{1/2}(F=I+1/2)$
$D_1$ transition at 817~nm and the trap laser to the 718~nm
$7S(F=I+1/2)-7P_{3/2}(F=I+3/2)$ cycling transition; for atom
cooling the second laser is slightly red-detuned. Tuning is very
critical: in order to reproducibly trap each Fr isotope, it is
important to know the laser absolute frequency with an uncertainty
comparable to the natural width ($\Gamma$=7.6~MHz) of the line.
However the best available measurement was performed only for
$^{210}$Fr with a wavemeter and was affected by a 90~MHz
uncertainty~\cite{orozco}. So we decided to implement a system
that allowed us to improve the information about the trapping
frequencies and the energy of the atomic levels at stake, and here
we present the principle of the experimental method.

The measurement accuracy that can be expected for the frequencies
of trap transitions in Fr is limited to a few MHz by the natural
width $\Gamma$ and in our experiment the precision is also
constrained by the modest signal-to-noise ratio allowed by the
small number of atoms. In these conditions it was reasonable to
choose the classical methods of optical interferometry, without
resorting to the complex technique of femtosecond frequency combs.

\begin{figure}[tbp]
\begin{center}
\includegraphics[width=8cm]{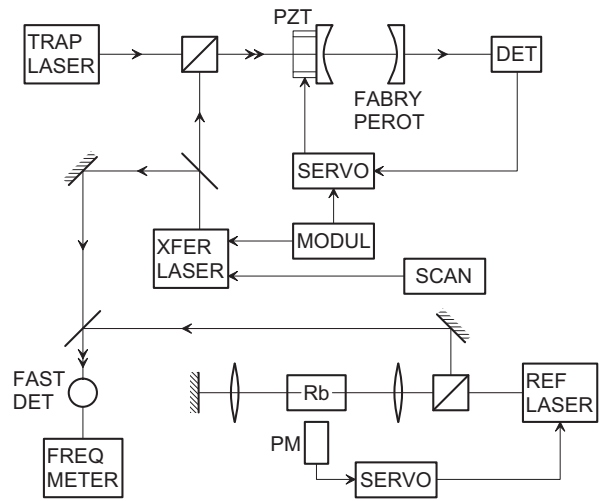}
\end{center}
\caption{\label{fig:alignment} Experimental apparatus. The trap
laser at 718~nm and the transfer laser operating near 778~nm are
carefully superimposed inside the Fabry-Perot interferometer. The
resonator length is driven by a piezo and it is locked to a
transmission peak of the transfer laser. This laser is manually
scanned, and the resonator follows it, in search of the
transmission peak of the trap laser through the interferometer.
The frequency of the transfer laser is measured by beating with
the reference laser. This secondary frequency-standard is a diode
laser locked to the 778~nm two-photon transition of $^{87}$Rb; PM
is a photomultiplier that collects the Rb blue fluorescence.}
\end{figure}

We then developed a methodology based on the use of a Fabry-Perot
resonator to measure the frequency $\nu _1$ of the Ti:Sa laser
with respect to a well-known frequency $\nu _0$ of a different
laser ($\nu _0$ is attained by comparison with a third laser, the
frequency standard). The principle of the experimental apparatus
is described in Fig.~\ref{fig:alignment}. The idea is to overlap
the two laser beams and observe the interferometer transmission:
for very accurate measurements the interferometer should be
operated in vacuum and at different mirror
spacings~\cite{garreau90}. Unfortunately, in the first
implementation of our setup we could not satisfy these
requirements, therefore much of the following discussion is
devoted to explain how we treated and evaluated the systematic
effects of the air refractive index and of the reflective phase
shift. Actually when both lasers are transmitted by the
Fabry-Perot, their frequencies are linked by the resonance
condition~\cite{kogelnik,garreau90}
\begin{equation} \label{eq:resonance}
    \nu _i \cdot n(\nu _i) = \frac{c}{2d} ~ \left( N_i-\frac{\psi (\nu
    _i)}{2 \pi} + \frac{\phi _i}{\pi}
    \right), ~~~i=0,1,
\end{equation}
where $n$ is the frequency-dependent refractive index of air, $c$
is the speed of light, $d$ the mirror separation and $N_i$ the
interference order (integer). $\psi (\nu _i)$ is the phase shift
due to the reflection on both mirrors in a round-trip while $\phi
_i$ is the Fresnel phase shift, which affects the laser beam in a
single pass between the mirrors: for a TEM$_{pq}$ mode in a cavity
with mirrors of equal curvature $R$, $ \phi _i = (p+q+1) ~ \arccos
(1-{d}/{R})$.

We first discuss the effects of the dispersion in the refractive
index of the air. We can determine the relation between the two
frequencies $\nu _0$ and $\nu _1$ imposed by the resonance
condition~(\ref{eq:resonance}) by taking the difference
\begin{equation} \label{eq:nu2nu1}
    \nu _1 \frac{n_1}{n_0} - \nu _0 = \frac{c}{2d~n_0} \left(\Delta N -
    \frac{\psi _1-\psi _0}{2 \pi} + \frac{\phi _1-\phi
    _0}{\pi}\right) ,
\end{equation}
where $n_i$, $\psi _i$ stand for $n(\nu _i)$, $\psi (\nu _i)$ and
$\Delta N = N_1 -N_0$.

For measurements in air we must know the refractive index $n(\nu
)$ in order to derive the frequency $\nu _1$ from
Eq.~(\ref{eq:nu2nu1}). $n(\nu )$ depends on several environmental
parameters: temperature, atmospheric pressure, humidity, CO$_2$
concentration. During each run, we estimated these quantities and
used the results to find $n(\nu)$, computed according to
Ref.~\cite{bonsch98}. The final precision for $n$, better than $2
\cdot 10^{-6}$, is limited essentially by the accuracy of the
temperature and pressure data ($< 2 ~^\circ$C and $< 3$~Torr
respectively). The contribution due to the uncertainties of the
humidity and of the CO$_2$ concentration were completely
negligible, and also the error (0.2~ppm) on the frequency $\nu$,
initially measured with a wavemeter, did not affect the accuracy
of $n(\nu )$.

It is worth noting that the error for the ratio $n_1/n_0$ is even
lower than the absolute accuracy for $n$: we found that for our
frequencies $\nu _0 \simeq 385$~THz and $\nu _1 \simeq 417$~THz,
it is better than $5 \cdot 10^{-9}$. This is the reason why we
expressed the result in Eq.~(\ref{eq:nu2nu1}) as a function of
$n_1/n_0$. We will show below that the factor depending on $n_0$
in the second member of that equation can be calibrated with a
dedicated procedure. In summary, the contribution of the
refractive index of air to the final error for the frequency
measurements was conservatively estimated to be less than 2~MHz.

Now we discuss how we managed to evaluate the uncertainty produced
by the reflective phase shift of the mirrors. We used a confocal
Fabry-Perot cavity ($d=R$). In this case, $\phi _i$ is a multiple
of $\pi /2$, therefore it is not strictly necessary to match the
laser beam to the TEM$_{00}$ mode of the cavity: simply, with both
$(p+q)$ even- and odd-order modes, the free spectral range (FSR)
appears to be $c/4d$ instead of $c/2d$. However, if we manage to
align the laser beam very well on the cavity axis, we observe that
the odd modes are strongly suppressed for symmetry reasons and we
recover the FSR~$=c/2d$. We took advantage of this property to
optimize the alignment of the laser beams in the cavity, by
minimizing the intensity of transmission peaks corresponding to
odd modes: this ensures that the travelled distance $d$ is the
same for both lasers. For even modes, we obtain that $(\phi _1 -
\phi _0)/\pi$ is integer, therefore
\begin{equation} \label{eq:Q}
    \nu _1 \frac{n_1}{n_0} - \nu _0 = \frac{c}{2d~n_0} \left(\mathcal{N} -
    \frac{\psi _1 - \psi _0}{2 \pi} \right),
\end{equation}
where the integer $\mathcal{N} = \Delta N + (\phi _1 - \phi
_0)/\pi$ in our apparatus is typically near 8000.

In general, the reflective phase shift $\psi$ is not zero and
depends on the frequency of the light. If we assume that the
dependence is linear in a frequency interval corresponding to the
high-reflectivity range of the mirrors~\cite{handbook,seeley:64},
then $\psi (\nu ') = \psi (\nu ) + 2 \pi \alpha (\nu ' - \nu )$.
The validity of the linear approximation is discussed in the
following. From Eq.~(\ref{eq:Q}), we find that
\begin{equation*}
    \nu _1 \frac{n_1}{n_0} - \nu _0 = \frac{c}{2d~n_0} \left[\mathcal{N} -
    \alpha (\nu _1 - \nu _0) \right],
\end{equation*}
and with a bland approximation,
\begin{equation} \label{eq:alpha}
    \nu _1 \frac{n_1}{n_0} - \nu _0 =
    \frac{\dfrac{c}{2d~n_0}}{1 + \dfrac{c}{2d~n_0} ~\alpha} ~ (\mathcal{N} + \epsilon ) \simeq \mathcal{F} \cdot
    \mathcal{N},
\end{equation}
where the factor $\mathcal{F}$, implicitly defined by the last
equate, has the meaning of an effective FSR. The error in the
approximation is $\epsilon = \alpha \, (n_1/n_0-1) \, \nu _1$: it
is expected to be negligible at our level of accuracy. To get an
idea of its order of magnitude, let's recall the operation of a
multi-layer dielectric quarter-wave stack: for a typical structure
of several alternating layers of high and low refraction, a
reasonable value for $\alpha$ is $-3 \cdot 10^{-15}$~Hz$^{-1}$.

We describe the calibration procedure by which we can determine
the factor $\mathcal{F}$, necessary to compute the unknown
frequency from Eq.~(\ref{eq:alpha}). The idea is to tune the Ti:Sa
laser to several different frequencies $\nu _1$ corresponding to
the transmission peaks of the Fabry-Perot while its length is
locked to the resonance value found for the trap frequency; the
scanned range is $350-430$~THz and we measure them with our
commercial wavemeter (10~MHz resolution, 0.2~ppm quoted accuracy).

For each measurement, we know that the condition (\ref{eq:alpha})
has to be satisfied, with an integer $\mathcal{N}$ corresponding
to the number of FSRs separating $\nu _1$ from $\nu _0$. In
practice, with frequencies $\nu _1$ near $\nu _0$, it is possible
to directly count the number of FSRs, find $\mathcal{N}$ and
obtain a first estimate of $\mathcal{F}$. This value, together
with the integer condition for $\mathcal{N}$ and the wavemeter
measurement, is used in Eq.~(\ref{eq:alpha}) to determine a new
$\mathcal{N}$ for a frequency $\nu _1$ farther from $\nu _0$, and
to find a more accurate value of $\mathcal{F}$. But more accurate
values of $\mathcal{F}$ allow us to extract $\mathcal{N}$ for
frequencies ever more distant: by iterating the procedure, we find
$\mathcal{N}$ for all the measured frequencies. We are then able
to fit the function
\[
    \nu _1 \frac{n_1}{n_0} = a + b \cdot \mathcal{N}
\]
to the whole set of $\nu _1 n_1 / n_0$ acquired values, to find
the parameters $a$ and $b$. $a$ is compared to the known value of
the transfer laser frequency $\nu _0$ to deduce the systematic
error of the wavemeter, which was found to be 50~MHz at $\nu _0
\simeq 385$~THz. This estimation allows us to introduce a small
correction to compensate for the bias of the wavemeter and
determine $\mathcal{F}$:
\[
    \mathcal{F} = b \cdot \frac{\nu _0}{a}.
\]

Note that the fitting procedure also allows us to check that
Eq.~(\ref{eq:alpha}) is satisfied in a large range of frequencies,
and therefore gives an experimental support to the assumption of
negligible non-linearity for the reflective phase shift in our
spectral region.

A crucial element for the evaluation of the final uncertainty is
the accuracy of the reference laser. As explained in
Fig.~\ref{fig:alignment} the calibrated frequency is provided by
the coordinated operation of the transfer laser and the reference
laser. They are diode lasers in an extended cavity configuration,
with a Littrow-mounted grating for wavelength selection  and
piezoelectric stacks for fine tuning. The frequency of the
reference laser is defined by a secondary frequency standard,
namely the $5S-5D_{5/2}$ two-photon transition in Rb atoms, at
778~nm~\cite{touahri97}. In order to implement this standard, we
lock the red emission of the laser by observing the blue
fluorescence at 420~nm from the $6P_{3/2}$
level~\cite{sanguinetti}. Usually, we excite the
$5S(F=2)-5D_{5/2}(F=4)$ two-photon hyperfine component of
$^{87}$Rb (nominally at $385284566.366 \pm 0.008$~MHz), because it
is relatively intense and well separated from the contiguous
hyperfine lines. A fast photo-diode followed by a radio-frequency
amplifier is used to detect the beat signal between the transfer
laser and the reference laser and its frequency is measured by a
counter: so the absolute frequency of the transfer laser can be
obtained.

In fact, to check the reliability of our frequency standard, we
set up two laser systems, independently locked to the two-photon
transition of Rb in separate cells, and compared their
frequencies. We measured the beat frequency of the two lasers in
different conditions: cell temperature, beam alignment, selected
hyperfine component. In particular we changed the focusing of the
laser beams inside the Rb cell, to observe the effects caused by
the light shift and we also measured the frequency shift as a
function of an external applied magnetic field, for testing the
efficiency of our mu-metal shield. In normal conditions, we can
safely estimate an absolute accuracy of the reference frequency
better than 300~kHz.

Our Fabry-Perot resonator has a nominal free spectral range
$c/4d=2$~GHz and a finesse 200 in the range $700-860$~nm. The
mechanical structure supporting the mirrors is a thick super-invar
cylinder and the spacing can be finely adjusted by means of a
piezo. At the beginning of each measurement run, both lasers are
carefully aligned according to the procedure described above.

As explained in Fig.~\ref{fig:alignment} we first obtain the
resonance condition for the transfer laser, then we search for the
transmission of the Ti:Sa laser. It was not necessary to separate
the two transmitted beams: we directly observe the sum of the two
signals with the same detector. The feedback loop is robust, so
for reasonable Ti:Sa intensities the cavity lock is not disturbed.

\begin{figure}[tbp]
\begin{center}
\includegraphics[width=70mm]{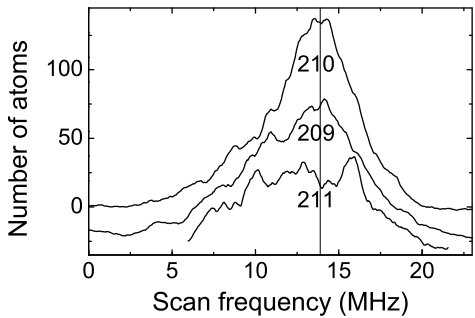}
\end{center}
\caption{\label{fig:results}Number of trapped atoms as a function
of the Ti:Sa laser scan frequency $\nu _1$, for the three isotopes
$^{209}$Fr, $^{210}$Fr, and $^{211}$Fr. Each scan begins with no
trapped atoms: two of the curves are down-shifted for clarity.}
\end{figure}

Once we have maximized the transmission for both lasers, we can
use Eq.~(\ref{eq:alpha}) and the performed calibration to find the
accurate value of the Ti:Sa laser frequency. By repeating the
measurement several times, we found that the reproducibility is
better than 2~MHz, mostly limited by the localization of the
transmission peak for the Ti:Sa laser. We measured the number of
trapped atoms for $^{209}$Fr, $^{210}$Fr, and $^{211}$Fr, as a
function of the trapping-laser frequency (Fig.~\ref{fig:results}).
Due to a limited Fr production at that run-time, we did not
operate in optimal conditions but the signal was high enough to
determine the frequency corresponding to the maximum number of
atoms. Then we measured the absolute frequencies corresponding to
the peak signal for the three isotopes. The data for the
calibration procedure (33 $\nu _1$ frequencies in total) were
acquired at the cavity lengths corresponding to the three trapping
frequencies and allowed to obtain the most accurate calibration of
the $\mathcal{F}$ factors ($\Delta \mathcal{F} < 400$~Hz), which
means a contribution of about 3~MHz for the error on the final
frequencies.

We report in Table~\ref{tab} the results of our measurements,
along with the frequencies of the repumping transitions, measured
with the wavemeter after calibration with the secondary frequency
standard. If we assume that the trapping laser detuning
corresponds to $5\Gamma \pm 2\Gamma$ (typical values for a MOT),
we can deduce the frequency of the atomic lines involved in the
trapping process (Table~\ref{tab}). In order to confirm the
correctness of this approach, we used it to measure the frequency
of the $5S(F=3)-5P_{3/2}(F=4)$ transition in a $^{85}$Rb MOT.
After applying the detuning correction, we measured $384229.250
(18)$~GHz, compatible with the most accurate value presently
available, namely $384229.2428 (4)$~GHz~\cite{barwood}.

In conclusion we presented accurate measurements of the
laser-trapping frequencies of $^{209}$Fr, $^{210}$Fr, and
$^{211}$Fr. Most of the results displayed in Table~\ref{tab} are
new;  in cases where previous measurements are available, our
results agree with existing data, but feature an improved
accuracy. The interferometric method, much simpler than other
spectroscopic techniques, allowed us to reach a precision of
7-9~MHz with samples of about 100 atoms. The experimental setup is
presently being upgraded by operating in vacuum and using the
virtual mirrors method~\cite{garreau90}, and we expect an
improvement of about a factor 2 in the final accuracy.

\begin{table}[tbp]
  \begin{tabular}{|c|c|c|c|} \hline
    Isotope & Trapping laser    & Repumper   & Trapping transition \\\hline
    209     & 417415.087(8)   & 366897.428(50)      & 417415.125(17) \\
    210     & 417412.448(7)   & 366898.698(50)      & 417412.486(17) \\
    211     & 417412.627(9)   & 366895.568(50)      & 417412.665(18) \\\hline
  \end{tabular}
\caption{\label{tab}Frequencies in GHz of the trapping and
repumping lasers that maximize the number of atoms in the MOT. We
also report the frequency corresponding to the atomic lines
involved in the trapping transition (corrected for the laser
detuning). The quoted errors are at 95\% confidence level.}
\end{table}

The importance of precision measurements on Fr, particularly if
available on several different isotopes, lies in the interest of
this heavy atom for testing fundamental symmetries: both parity
violation and EDM require a theoretical analysis of the atomic
structure that must be checked against experimental data. Accurate
absolute measurements are also very useful in the design of future
experiments. Therefore the results presented here are an important
step in this direction.

\end{document}